\documentstyle[aps,amssymb,multicol]{revtex}

\include{iopart12}
\begin{document}
\tightenlines
\def\CC{{\rm\kern.24em \vrule width.04em height1.46ex depth-.07ex
\kern-.30em C}}
\def\P{{\rm I\kern-.25em P}}
\def\RR{{\rm
         \vrule width.04em height1.58ex depth-.0ex
         \kern-.04em R}}
\def\id{{\rm 1\kern-.22em l}}

\title{Bethe Ansatz  solution of a new class of Hubbard-type models}
	
\author{Andreas Osterloh$^{1}$, Luigi Amico$^{2,3}$, and 
Ulrich Eckern$^{1}$}
\address{$^{1}$Institut f\"ur Physik,Universit\"at Augsburg, D-86135 Augsburg, Germany}
\address{$^{2}$ Departamento de F\'{\i}sica Te\'orica
de la Materia Condensada $\&$ Instituto ``Nicol\'as Cabrera''
\\
Universidad Aut\'onoma de Madrid, E-28049 Madrid, Spain.}
\address{$^{3}$ Dipartimento di Metodologie Fisiche e Chimiche
per l'Ingegneria, Facolt\'a di Ingegneria,
Universit\'a di Catania \& INFM, viale A. Doria 6, I-95129 Catania,
Italy.}

\maketitle
\begin{abstract}
We define one-dimensional particles 
with generalized exchange statistics.
The exact solution of a Hubbard-type Hamiltonian   
constructed with such particles is achieved using the Coordinate Bethe Ansatz.
The chosen deformation of the statistics is equivalent to the presence of a magnetic field
produced by the particles themselves, which is present also in a ``free gas''
of these particles.
\end{abstract}
\vspace{-1.2cm}
\begin{multicols}{2}
\narrowtext
Studies on strongly correlated one-dimensional (1D) systems have deeply influenced 
modern concepts in many-particle physics. 
These systems may be of strictly linear extension like
quantum wires, quantum-Hall bars, (quasi-)1D organic metals and spin chains, or 
higher dimensional systems where  rotational degrees of freedom are not relevant, 
as in the Kondo problem~\cite{1D-EXP}. Exactly solvable models~\cite{MATTIS} 
play a crucial role in the theoretical description of 1D systems, since there, most of the 
results obtained by approximate methods are not 
reliable (especially at low temperature). They also serve as 
a reliability test for various approximations, which are applied to
models in higher dimensions.
\\
An important example is the
Bethe Ansatz (BA) solvable 1D-Hubbard model (HM)~\cite{LIEBWU}.
The HM, originally proposed in Ref.~\onlinecite{HUBBARD}, 
describes electrons hopping on a $D$-dimensional lattice, while
experiencing an on-site interaction.
The two-dimensional HM  is believed to capture important features 
of high-$T_{c}$ superconductivity\cite{HIGHTC}.      
Away from half filling, the 1D-HM exhibits Luttinger Liquid (LL) behavior~\cite{LUTTINGERLIQ}.

Various generalizations of the HM have been proposed;
most of them emerge from changing explicitly the form of the Hamiltonian~\cite{MARTINS}.
The coupling of the fermionic degrees of freedom 
to a phononic bath has been studied in Ref.~\cite{MORA}. 
Recently, Schulz and Shastry\cite{SCHULZ} 
proposed a class of solvable Hubbard-type models in which the hopping 
amplitude of the particles with a given spin orientation is modulated by a gauge field 
(entering the Hamiltonian via a Peierls-like substitution),
which depends on the density of particles with opposite spin orientation.  
A similar gauge-like potential has been employed to couple 
several Hubbard chains\cite{ZVYAGIN}, where 
``$M$-colored'' electrons in $M$ chains perform
intra-chain hopping, which is (gauge-like) affected by the density of electrons 
in the other chains~\cite{ZVYAGIN}.
The effective inter-chain interaction resembles an electromagnetic 
field due to the motion of electrons moving along each chain.
The models studied in Ref.\cite{SCHULZ} show LL behavior even in the absence 
of the interaction: the asymptotic behavior of correlation functions depends on 
the coupling strength between the gauge field and the particles, which is responsible for the 
change in periodicity of the wave function. 
The gauge potential   breaks the time reversal symmetry of the 
Hamiltonian. Charge and spin excitations 
have anyon-like character, and the scaling properties deviate from the 
``fermionic'' LL theory~\cite{SCHULZ}.   
\\
Another way of extending a model, is to modify the ``nature'' of the particles  
entering the Hamiltonian, without changing its formal structure\cite{MAASSARANI}. 
In the present paper we choose this approach to extend the HM: 
we keep the formal structure of the Hubbard Hamiltonian unaltered, but we change the
particles from fermions to particles whose wave functions obey a generalized exchange symmetry.
We call the statistics of such particles Deformed Exchange Statistics (DES)~\cite{ANY}.
Here, a variant  of the DES introduced in Ref.\cite{AMOSECK} is used.
The corresponding particles constitute a non-Abelian realization of the symmetric group $S_N$ 
(see Ref.~\cite{ANY}).
The ``deformed'' 1D HM is solved exactly by means of the Coordinate BA (CBA) and 
the Bethe Equations (BE) are compared with the results in Refs.~\cite{SCHULZ} 
and~\cite{SHASTRY}. 

We define the HM Hamiltonian for particles obeying DES as  
\begin{equation}\label{ourHAMILTON}
H=  - {t}\, \sum_{i,\sigma } 
\, \left ( f_{{ i},\sigma}^\dagger f_{{i+1},\sigma} + 
{\rm h.c.} \right ) +
U\, \sum_{ i}\nu_{{i},\uparrow} \nu_{{ i},
\downarrow} \; , 
\label{MONRAMODEL}
\end{equation}
where $f_{i,\sigma}$, $f_{i,\sigma}^{\dagger}$ 
($\sigma \in \{\uparrow,\downarrow \}$, respectively $\sigma\in \{1/2,-1/2 \}$) 
obey the  deformed relations
\begin{equation}
\begin{array}{rcl}
f_{j,\sigma}^\dagger f_{{ k},\sigma'} + {\cal Q}_{j,k}^{\sigma,\sigma' }\,  
f_{{ k},\sigma'} f_{{ j},\sigma}^\dagger &=& \delta_{j,k}\ \delta_{\sigma \, \sigma '} ,\\
f_{{ j},\sigma} f_{{ k},\sigma'} + {\cal Q}_{k,j}^{\sigma',\sigma} \,  
f_{{ k},\sigma'} f_{{ j},\sigma} &=& 0 ,
\end{array}
\label{any}  
\end{equation}
\vspace{-0.4cm}
\begin{equation}
{\cal Q}_{j,k}^{\sigma, \sigma'} = ({\cal Q}_{k,j}^{\sigma',\sigma})^{-1}  = 
{{\cal Q}^{\dagger}}_{k,j}^{\sigma', \sigma} \; ,
\label{CONS1}
\end{equation}
\vspace{-0.8cm}
\begin{equation}
[ f^\dagger_{i,\sigma}\, ,\, {\cal Q}_{j,k}^{\sigma,\sigma'}] \ 
=\ [ f_{i,\sigma} \, ,\, {\cal Q}_{j,k}^{\sigma,\sigma'} ] \ = \ 0 \; 
\label{postulate-rel} 
\end{equation}
Relations (\ref{any}) are formally analog to quon commutation 
rules~\cite{WU}, but here the deformation parameter depends on  indices $(j,\sigma| k, \sigma')$. 
The operators $\nu_{{ j},\sigma}\doteq 
f_{{ j},\sigma}^\dagger f_{{ j},\sigma}$ are the particle-number 
operators. 
Equation (\ref{CONS1}) guarantees that the particles are representations 
of $S_N$, whereas Eq. (\ref{postulate-rel}) ensures the standard commutation relations 
\(
[ \nu_{i,\sigma},  \nu_{j,\sigma'} ]=0
\), 
\(
[ \nu_{i,\sigma}, f_{j,\sigma'}^\dagger ]=
\delta_{i,j}\delta_{\sigma,\sigma' }f_{j,\sigma'}^\dagger\), and  
\(
[ \nu_{{i},\sigma}, f_{{ j},\sigma'}]=-
\delta_{i,j}\delta_{\sigma,\sigma' }f_{{ j},\sigma'}\),
providing a well defined Fock representation of the algebra defined in Eqs.~(\ref{any}).
We point out that DES is defined for any ${\cal Q}_{j,k}^{\sigma,\sigma' }$ fulfilling 
(\ref{CONS1}), (\ref{postulate-rel}); we choose the deformation parameter depending on spins and coordinates,
but also such that the solvability of the model (\ref{MONRAMODEL}) is ensured:
\begin{eqnarray}\label{OURQ}
&{\cal Q}_{j,k}^{\sigma,\sigma'}:=\left \{ \begin{array}{ll} e^{i\psi(\sigma+\sigma')}
e^{i\tilde{\Phi}\cdot(\sigma-\sigma')},\ \quad & j>k,  \\
e^{i\Phi\cdot(\sigma-\sigma')}, \quad & j=k, \\
e^{-i\psi(\sigma+\sigma')} e^{i\tilde{\Phi}\cdot(\sigma-\sigma')}, 
\quad & j<k.  \end{array} \right.  \quad &
\end{eqnarray}
For fixed $j,k,\sigma,\sigma'$, ${\cal Q}_{j,k}^{\sigma,\sigma'}$ is a $\CC$-number.

The deformation defined above becomes integrable if 
$\Phi-\tilde{\Phi}\pm\psi(0)=0\; {\rm mod}\; 2\pi$ holds.
This implies
$(\Phi-\tilde{\Phi};\psi(0))=(0;0)\ \vee\ (\pi;\pi)\ {\rm mod}\; 2\pi$.
We note that inserting definition~(\ref{OURQ}) in Eq.~(\ref{any}) 
implies the Pauli principle 
since ${\cal Q}_{j,j}^{\sigma,\sigma}=1$, and that two particles with different 
spin on the same site obey deformed  anticommutation relations since 
${\cal Q}_{j,j}^{\uparrow,\downarrow}\neq 1$.
Equation~(\ref{postulate-rel}) is trivially fulfilled ($q$ is a 
$\CC$-number) as well as the consistency relation (\ref{CONS1}) together 
with Eq.~(\ref{OURQ}).
The fermionic case is obtained setting
$\Phi=\tilde{\Phi}=0;\ \psi\equiv 0$, whereas the hard-core bosonic case is 
covered via $\Phi=\tilde{\Phi}=0;\ \psi(\pm1)\equiv \pi$.
Furthermore we mention that for spin-$1/2$ particles and factorizing
spin- and coordinate dependence of the deformation, every spin dependence 
in the deformation can be written in the linear form used here.

It is worth mentioning that (\ref{OURQ}) implies fixing an order
on the ring. Such an order distinguishes
one (given) lattice site from the others in that it defines the ``beginning'' 
of the ring, where one starts counting. Periodicity means not only that
the wave function has to be periodic, but simultaneously demands the result
being independent of where the starting point was set.
What is equivalent, is fixing a period 
$P_0 \doteq \{\,j_1\, ,\,  \dots\, ,\, j_1 + L\, \}$ on the infinite 
periodic chain. Consistency of the Periodic Boundary Condition  
(see below)  with this induced order is given if the results are 
independent of $P_0$. In the sequel it will be seen that this condition 
is fulfilled. 

Now we show that the Hamiltonian~(\ref{MONRAMODEL}) is exactly solvable by means 
of CBA. We apply Periodic Boundary Conditions (PBC), which means
$f_{{ j+L},\sigma}\equiv f_{{ j},\sigma}$.
The general $N$-particle state in a chain with $L$ sites can be written as 
\begin{equation}
|\Psi \rangle = \sum_{\pi \in {S_N}} \; \sum_{j_1 .. j_N}
 \Psi(\vec{j_{\id}}\;|\pi )\;  {\cal F}^\dagger (\vec{j_{\id}}\;|\pi) |0 \rangle ,
\end{equation}
where  we have used the abbreviations
$\vec{j_{\gamma}}\doteq (j_{\gamma(1)},\dots,j_{\gamma(N)})$, 
${\cal F}^\dagger (\vec{j_{\gamma}}\;|\pi) 
\doteq f_{j_{\gamma(1)},\sigma_{\pi(1)}}^{\, \dagger}\dots
f_{j_{\gamma (N)},\sigma_{\pi(N)}}^{\, \dagger}$, for arbitrary $\gamma \in S_N$; 
$\id$ is the identity in $S_N$.
For indistinguishable particles in $D\neq 2$, a Fock base state
is uniquely determined by the coordinates and the spin-configuration; 
it must not depend on the order in which the particles are created.
This requires
\begin{equation}
\Psi(\vec{j_{\gamma}}\;|\gamma\circ\pi)
{\cal F}^\dagger (\vec{j_{\gamma}}\;|\gamma\circ\pi)
\equiv 
\Psi(\vec{j_{\gamma'}}\;|\gamma'\circ\pi)
{\cal F}^\dagger (\vec{j_{\gamma'}}\;|\gamma' \circ\pi)
\label{WELLDEF}
\end{equation}
for any $\gamma , \gamma'\in S_N$, since changing the order in creating particles means
permuting both the coordinate and spin indices of the operators 
$f_{j_{\gamma (l)},\sigma_{\pi(l)}}^{\, \dagger}$ in ${\cal F}^\dagger$.   
Equation~(\ref{WELLDEF}) fixes the symmetry of the wave function $\Psi$. 
In the case of fermionic statistics, $\Psi(\vec{j_{\gamma}}\;|\gamma\circ\pi)
= {\rm sign}(\gamma)\Psi(\vec{j}_\id\;|\pi)$, where ${\rm sign}(\gamma)=+/-$ for even/odd 
permutations.
Every permutation can be decomposed into a product of neighbor exchanges
 $\chi_k\doteq k\leftrightarrow (k+1)$. Taking
$\gamma=\chi_n\circ\dots\circ\chi_2\circ\chi_1$ as a simple and relevant 
example to fix the idea and further defining the partial products as
$\gamma_l\doteq \chi_l\circ\dots\circ\chi_2\circ\chi_1$ for $l\leq n$, we obtain
\begin{equation}\label{GENERALQ}
\Psi(\vec{j}_\gamma \; |\gamma\circ\pi)
=\prod_{l=1}^n\left [- {\cal Q}_{x'(l),x(l)}^{s'(l),s(l)}\right ]\Psi(\vec{j}_\id \; |\pi),
\end{equation}
where $x(l)=j_{\gamma_l\circ\pi(l)}$, $s(l)=\sigma_{\gamma_l\circ\pi(l)}$, 
$x'(l)=j_{\gamma_l\circ\pi(l+1)}$, and $s'(l)=\sigma_{\gamma_l\circ\pi(l+1)}$. 

Due to the DES~(\ref{any}), special care must be taken when
employing the symmetry of $\Psi$  to restrict the coordinate space to a
sector of ordered coordinates (called ``order region'' in the following).
An order region can be characterized by a permutation $\tau \in S_N$ which connects 
the actual  set of coordinates $(j_1,\dots,j_N)$ with the ordered one so 
that $\left ( j_1,\dots,j_N\right ) \doteq ( i_{\tau(1)},\dots,i_{\tau(N)} )$, where 
$i_1 \leq \dots \leq i_n$. 
The corresponding wave function, which is defined in this order region, is labeled 
by $\tau$: $\Psi_\tau(\vec{j}_{\pi'}\; |\pi)\doteq \Psi_\id(\vec{j}_{\tau\circ\pi'}|\tau\circ\pi)$. 
At first, it is necessary keeping this label for writing the Schr\"odinger Equation, 
since the hopping term can connect different order regions. 
Due to the Pauli principle, only two order regions are needed to solve
the Schr\"odinger Equation (let these two order regions be connected by $\chi_k $). 
For $\gamma=\chi_k $, Eq.~(\ref{GENERALQ}) simplifies to 
\begin{equation}\label{SYMM-DIST}
\Psi_{\chi_k\circ\tau}(\vec{i}\;|\pi) = 
       -{\cal Q}_{x,x'}^{\sigma,\sigma'}\Psi_{\tau}(\vec{i}\;|\pi)  ,
\end{equation}
where $x=i_{\pi(k)}$, $\sigma\doteq\sigma_{\pi{\scriptscriptstyle(k)}}$ and  
$x'=i_{\pi(k+1)}$, $\sigma'\doteq\sigma_{\pi{\scriptscriptstyle(k+1)}}$.
In the special case of a doubly occupied site, $i_k = i_{k+1}$, the symmetry relation
on the border of an order region leads to
\begin{equation}\label{SYMM-ONSITE}	
\Psi_{\chi_k}(\vec{i}|\pi) = \Psi_\id(\vec{i}\;|\chi_k\circ\pi) = 
      - {\cal Q}_{x,x}^{\sigma,\sigma'}\Psi_\id(\vec{i}\;|\pi),
\end{equation} 
Equations (\ref{SYMM-DIST}) and (\ref{SYMM-ONSITE}) allow writing the 
Schr\"odinger Equation in terms of a single $\Psi_\tau$~\cite{SYMM-EQS}. 
Since each $\Psi_\tau$ can be obtained from
$\Psi_\id$ using Eq. (\ref{GENERALQ}), we consider $\Psi_\id$ in the following. 
The label $\id$ will be omitted, and the coordinate vector $\vec{j}$ is chosen 
being ordered.
We insert the Bethe wave function 
$$\Psi(\vec{j}\;|\pi)=\sum\limits_{{\pi'} \in {S_N}} A({\pi'}|\pi)
\exp{\left [i\sum\limits_{m=1}^N j_{m}p_{{\pi'}(m)}\right ]}$$ 
into $H|\Psi\rangle=E|\Psi\rangle$.
For pairwise distinct coordinates we obtain the energy $E$ in terms of the momenta $p_l$: 
$
E=-2t\sum_{l=1}^N \cos{p_l} .
$
Its form 
is unaltered
by the deformation parameter $q$.
Using Eqs. (\ref{SYMM-DIST}) and (\ref{SYMM-ONSITE}), 
the Schr\"odinger Equation corresponding 
to doubly occupied sites explicitly contains 
the deformation parameter, and the scattering matrix $S$ reads
\begin{equation}\label{S-matrix}
S(\lambda,\lambda';m)=- \frac{
     i(\lambda-\lambda')e^{i\Phi\cdot(\mu-\mu')}\P_{m,m+1} 
       -\frac{U}{2t}}{
     i(\lambda-\lambda')-\frac{U}{2t}}, 
\end{equation}
where we have defined  $\lambda\doteq\sin (p_{{\pi'}(m)})$, 
$\lambda'\doteq\sin (p_{{\pi'}(m+1)})$, 
$\mu\doteq\sigma_{\pi(m)}$, and $\mu'\doteq\sigma_{\pi(m+1)}$. 
The permutation operator $\P_{m,m+1}$ is defined by its action on 
the amplitudes $A(.|.)$:
$\P_{m,m+1}\,A({\pi'}|\pi)\doteq A({\pi'}|\chi_m\circ\pi)$.
The scattering matrix $S$ fulfills the Yang-Baxter Equation (YBE)
\begin{eqnarray}
S(\lambda,\lambda';m+1)\; && S(\lambda,\lambda'';m)\; S(\lambda',\lambda'';m+1)=  \\
&&S(\lambda',\lambda'';m)\; S(\lambda,\lambda'';m+1)\; S(\lambda,\lambda';m), \nonumber 
\end{eqnarray}
where $\lambda''\doteq \sin (p_{{\pi'}(m+2)})$.
For the validity of the YBE, it is important to notice that $\P$ and 
$(\mu-\mu')$ 
do not commute, since $\mu$ and $\mu'$ contain the spin permutation, 
which is affected by $\P$:
$\P_{m,m+1} (\mu-\mu') = (\mu'-\mu) \P_{m,m+1}$.
Applying PBC, which also demands independence from the pre-chosen order, 
we obtain
\begin{equation}\label{jPBC}
\Psi(j_1,..,j_{N}|\pi)=Q_{PBC}
\Psi(j_{k+1},..,j_N,j_1+L,..,j_k+L|\tilde{\pi}),
\end{equation}
where $\tilde{\pi}$ is the cyclic permutation 
$\tilde{\pi}\doteq {\,1\,\dots\,N-1\,N \choose k+1\,\dots\,k-1\,k}$, and the 
function $Q_{PBC}$ reads
\begin{eqnarray}\label{Q-PBC}
Q_{PBC}&=&(-1)^{k(N-1)}\prod_{m=1}^k\prod_{l=1}^{N-1}
           {\cal Q}_{i_{m+l},i_m}^{\sigma_{\pi(m+l)},\sigma_{\pi(m)}} \nonumber \\
&\equiv & (-1)^{k(N-1)} Q_{B,0}^k \prod_{m=1}^k Q_{B,\sigma_{\pi(m)}},
\end{eqnarray} 
where 
$Q_{B,\sigma}\doteq e^{i[(N_\sigma -1)\psi(2\sigma)-2\sigma\Phi N_{-\sigma}]}$ 
is the spin dependent part, 
and $Q_{B,0}\doteq 1$ the spin independent part of $Q_{PBC}$~\cite{STAT-Q}.
Thus, the deformation affects the BE 
in two distinct ways: (a) via the prefactor $Q_{PBC}$ by imposing PBC; its form arises  
from the deformed symmetry of the wave function induced by the algebra~(\ref{any}); 
(b) due to a modified $S$-matrix in case of on-site deformation.
The BE are
\begin{eqnarray}\label{BETHEEQ}
&&e^{i p_j L} = \frac{Q_S^{N_\downarrow}}{Q_{B,\uparrow} Q_{B,0}}  
                \prod_{a=1}^{N_\downarrow}
                \frac{i(\sin p_j - \zeta_a) -\frac{U}{4t}}
                        {i(\sin p_j - \zeta_a) +\frac{U}{4t}}, \\  
&&\prod_{b=1 \atop b\neq a}^{N_\downarrow}
\frac{i(\zeta_a - \zeta_b) +\frac{U}{2t}}{i(\zeta_a - \zeta_b) -\frac{U}{2t}}
= \frac{Q_{B,\downarrow} Q_S^N}{Q_{B,\uparrow}}    
          \prod_{l=1}^N
                \frac{i(\sin p_l - \zeta_a) -\frac{U}{4t}}
                        {i(\sin p_l - \zeta_a) +\frac{U}{4t}}, \nonumber   
\end{eqnarray}
where the factor $Q_S \doteq  e^{-i\Phi}$ arises from the $S$ matrix in 
diagonalizing of the transfer  matrix of the ``spin problem'', Eq.~(\ref{jPBC}). 
Using the definitions of $Q_{B,\sigma},Q_{B,0}$, $Q_S$, and the integrability 
condition $\Phi-\tilde{\Phi}\pm\psi(0)=0\; {\rm mod}\; 2\pi$, 
the factors due to the DES are
\begin{eqnarray}
\frac{Q_S^{N_\downarrow}}{Q_{B,\uparrow} Q_{B,0}} &=&
e^{-i(N_\uparrow -1)\psi(1)} \\
\frac{Q_{B,\downarrow} Q_S^N}{Q_{B,\uparrow}} &=&
e^{i\left[(N_\downarrow -1)\psi(-1)-(N_\uparrow -1)\psi(1)\right ]}.
\end{eqnarray}
We see how the above contributions involve only the spatial part of the 
deformation parameter; a purely spin-dependent deformation of the statistics
does not affect the spectrum of the model. Even in this case, however,  the eigenstates are 
different due to the modification in the $S$ matrix.
This can also be made transparent by representing the $f,f^\dagger$ by fermionic
operators $c,c^\dagger$:
$
f_{j,\sigma}\doteq c_{j,\sigma}\prod_{i=1}^L e^{-i\sigma\Phi n_{i,\sigma}}
$.
In a forthcoming paper~\cite{OAE} we will describe in more detail
the connection between fermionization and DES, as well as the connections to
Refs.~\cite{SCHULZ,SHASTRY}. Here we mention only that, defining
$\Phi_\sigma \doteq -(N_\sigma -1)\psi(2\sigma)$,
the phases appearing in the BE's decompose into 
$\Phi_\uparrow$ and $\Phi_\uparrow - \Phi_\downarrow$, found in Ref.~\cite{SHASTRY}. 
In contrast to the result in \cite{SCHULZ}, the factors $N_\sigma -1$ appear.
The reason for this is, that the phases in Ref.~\cite{SCHULZ} were produced by 
particles having opposite spin, whereas here the relevant phase comes from
particles having equal spin projections. Thus the particle feeling that phase
has to be excluded.

From the BE's (\ref{BETHEEQ}), the total momentum can be extracted:
\begin{equation}
P\doteq\sum_{i=1}^N p_j = -\sum_\sigma N_\sigma(N_\sigma -1)\frac{\psi(2\sigma)}{L}
= \sum_\sigma N_\sigma \frac{\Phi_\sigma}{L}.
\end{equation}
From this it is seen that the total momentum, though periodicity is implied,
is not necessarily a multiple of $2\pi/L$, which is the case for the undeformed 
model. This is not surprising, since the spectrum is that of a twisted fermionic model.
The total momentum can be obtained in shifting $N_\uparrow$ momenta of the undeformed 
model, which become $2\pi n/L$ for $U\rightarrow 0$, by $-(N_\uparrow -1)\psi(1)$
and the remaining $N_\downarrow$ momenta by $-(N_\downarrow -1)\psi(-1)$.
That meant
\begin{equation}
E=-2t\sum_\sigma\sum_{i_\sigma=1}^{N_\sigma} \cos(\frac{2\pi}{L} l_{i_\sigma}+
\frac{\Phi_\sigma}{L}),
\label{PHYSICAL}
\end{equation}
with $\Phi_\sigma$ as defined above in comparison with Ref.~\cite{SHASTRY}
and $l_{i_\sigma}$, for fixed $\sigma$, being distinct integers modulo $L$.
Expression (\ref{PHYSICAL}) shows 
the physical meaning of DES: The  deformation of the particles' statistics is 
equivalent to a magnetic field generated from the particles themselves. 
Such a magnetic field 
depends on how many particles are in the system. In particular (\ref{PHYSICAL}) 
shows how the energy of a gas of free ($U\rightarrow 0$) particles is not simply the sum 
of the single particles' energies, but it describes a system of interacting particles.
Such interaction purely comes from the deformed statistics.   
However, the noninteracting limit is to be taken carefully here.
This can be understood by facing the second BE. In order to keep the phase, 
two or more spin rapidities are forced to either coincide with each other
or with one or several $\sin (p_j)$. Such contributions have to be linear
in $U/4t$.

In conclusion, we have extended the Bethe ansatz for particles with deformed exchange
statistics or, which is equivalent, for solutions with generalized exchange symmetry.
A method already introduced in Ref.~\cite{AMOSECK} is here generalized for
problems including inner degrees of freedom like spin.
Using this technique, a new class of generalized Hubbard models could be shown
being exactly solvable. The spectrum equals that of a fermionic model
with spin dependent boundary phases, which in fact turns out being a general
feature of integrable deformed exchange statistics. 
The eigenfunctions however are different due to the non-fermionic exchange symmetry.
A preliminary study showed that these phases in general already 
contribute to the thermodynamic limit. We will discuss the details elsewhere~\cite{OAE}.
\\
A systematic study of deformed models is most interesting for at least two reasons.
First, ``integrable deformed statistics'' can be an alternative way of handling complicated 
interactions of integer statistics' particles. 
In a forthcoming paper~\cite{OAE} we will review these results from another view-point; 
namely fermionic models with correlated hopping. There, a general statement on
integrability of deformed exchange statistics will be drawn as well as for
models with correlated hopping.
\\
Second, it might be relevant
for modelling the edge states in fractional quantum Hall effect physics, the  
chiral Luttinger liquid behavior of which has recently been questioned~\cite{EDGE}.

\vspace{0.3cm}

We thank M. Rasetti 
for suggesting this line of research, and
R. Fazio, G. Falci, G. Giaquinta, S. Girvin, S. Isakov, A. Kundu, J. Myrheim,
A. Polychronakos and S. Sharov for fruitful discussions. 
L.A. acknowledges financial support from ``Fondazione A. della Riccia'' 
and the EU TMR Programme (ERB 4061 PL 95-0670), and the warm hospitality of 
the Theoretical Physics II group in Augsburg.
Support through the {\em Graduiertenkolleg} 
``Nonlinear Problems in Analysis, Geometry, and Physics" (GRK 283),
financed by the German Science Foundation (DFG) and the State of Bavaria, is acknowledged.

\end{multicols}


\begin{thebibliography}{100}
\bibitem{SYMM-EQS} Eqs. (\ref{SYMM-DIST}) and (\ref{SYMM-ONSITE}) guarantee that two wave 
	functions defined in neighboring regions 
	are single valued  on the intersection of their domains.
\bibitem{STAT-Q} The choice of the functional form of $q$, which was 
        integrable in~\cite{AMOSECK}, would destroy 
        the CBA  solvability of the  HM, since $Q_{PBC}$
        would become configuration dependent.
\end{thebibliography}
\end{document}